\documentclass[journal, twocolum, twoside]{IEEEtran}

\usepackage{amsmath}
\usepackage[dvips]{graphicx}
\usepackage{cite}
\usepackage{amssymb}
\usepackage{wasysym}
\usepackage{balance}
\usepackage{footnote}

\title{Partitioning of Distributed MIMO Systems based on Overhead Considerations}

\author{Athanasios~S.~Lioumpas, Petros~S.~Bithas,~\IEEEmembership{Member,~IEEE,}
         and Angeliki Alexiou,~\IEEEmembership{Member,~IEEE \vspace{-0.5cm}}
\thanks{This research has been co-financed by the European Union and Greek national funds through the Research Funding Program: Thales.}%
}

\begin{document}

\markboth{IEEE Wireless Communications Letters,~Vol.~X,
No.~XX,~XXXXX~2013}{Lioumpas, \MakeLowercase {\textit{et al.}}: Partitioning of Distributed MIMO Systems based on Overhead Considerations}

\maketitle

\begin{abstract}
Distributed-Multiple Input Multiple Output (D-MIMO) networks is a promising enabler to address the challenges of high traffic demand in future wireless networks. A limiting factor that is directly related to the performance of these systems is the overhead signaling required for distributing data and control information among the network elements. In this paper, the concept of orthogonal partitioning is extended to D-MIMO networks employing joint multi-user beamforming, aiming to maximize the effective sum-rate, i.e., the actual transmitted information data. Furthermore, in order to comply with practical requirements, the overhead subframe size is considered to be constrained. In this context, a novel formulation of constrained orthogonal partitioning is introduced as an elegant Knapsack optimization problem, which allows the derivation of quick and accurate solutions. Several numerical results give insight into the capabilities of D-MIMO networks and the actual sum-rate scaling under overhead constraints.
\end{abstract}

\begin{keywords}
Distributed MIMO, effective sum-rate, Knapsack optimization, network partitioning, overhead reduction.
\end{keywords}

\section{Introduction}\label{Sec:Introduction}
Distributed-Multiple Input Multiple Output (D-MIMO) networks (also known
as network MIMO) have attracted great research interest for their
potential to satisfy very high data rates requirements of future wireless networks \cite{J:NW_MIMO_3}. The generic system model comprises a number of
distributed access points (AP)s, communicating with a number of clients, forming a virtual MIMO array. Depending on the ratio between the number of APs and clients, as well as the kind of information that is shared among the network elements, several different techniques have been proposed, which aim at the interference mitigation, including interference alignment (IA), dirty paper coding and joint multi-user beamforming (JMB), each with different performance in terms of sum-rate \cite{aziz2012system}.

The data overhead significantly increases with the number of APs and clients, due to the substantial amount of information that must be shared among
the network elements for performing various operations, e.g., channel state information (CSI) estimation, time/frequency synchronization, data sharing,
\cite{J:heath_network_partitioning,C:Caire_MIMO}. In this sense, the effective sum-rate, i.e., the volume of the actual transmitted information bits, is reduced by a non-negligible factor as compared to the information-theoretic sum-rate \cite{J:heath_network_partitioning}. In this context, the overhead reduction of D-MIMO networks gained much research interest, with the efforts focusing on the selection (scheduling) of the APs involved in the network MIMO, \cite{C:Boccardi_MIMO}, \cite{J:caire_network_MIMO}, the reduction of CSI exchanges among the network \cite{ C:katabi_conf} and inter-cluster interference mitigation techniques \cite{J:Inter_cluster}.

More recently a novel concept has been introduced, where
a D-MIMO network employing IA is partitioned into smaller orthogonal D-MIMO groups (i.e., in a time division multiple access (TDMA) fashion) eliminating in this way any kind of interference \cite{J:heath_network_partitioning}. It was shown that when IA is employed in the full network, the effective sum-rate diminishes as the number of users increases, whilst the performance improves when orthogonal partitioning is employed. The partitioning algorithms in \cite{J:heath_network_partitioning} target at the maximization of the effective sum-rate, assuming that the size of the overhead subframe within the entire frame can change dynamically. Although, this is the optimal strategy in terms of sum-rate maximization, it may not be the case for several practical systems, where only a predetermined portion of the frame is available (e.g., in Long Term Evolution networks) \cite{B:Sengar_overhead}.

In this paper, we apply the partitioning concept to D-MIMO networks employing JMB, which does not require the network channel matrices to be known at both the transmitter and receiver sides. Furthermore, extending the scenario where the overhead size within the frame is unconstrained, we consider for the first time the case where the overhead size is constrained, in order to comply with potential requirements of practical telecommunication systems, e.g., \cite{B:Sengar_overhead}, \cite{C:LTE}. In this context, the partitioning optimization problem is formulated as an elegant Knapsack problem and its exact solution is computed. Our motivation is to provide the network designer the opportunity to compare the optimal performance of an unconstrained D-MIMO network, with the performance of a D-MIMO network where the overhead subframe size is constrained by the system frame structure.

\section{System Model}\label{Sec:system}
The downlink of a D-MIMO network is considered, where $K$ distributed APs (transmitters) communicate with $M=K$ spatially distributed single-antenna clients (users) through a time-varying fading channel. In this context and towards an interference-free communication scenario, the JMB concept is employed, \cite{C:katabi_conf}, assuming accurate CSI available at the transmitters as well as a high capacity backhaul link, similar to \cite{J:heath_network_partitioning,J:caire_network_MIMO,J:Inter_cluster}. Let $s_k, {\bf h_k}, {\bf w_k}, P_k$ denote the data symbol, the channel gain (amplitudes and phases) vector (complex row), the beamforming column vector and the transmit power allocated to user $k$. In this context, the beamforming weights $\left( {\bf W=[w_1, w_2, \ldots, w_K]}\right)$ are appropriately selected in order to satisfy the zero-interference condition, i.e., ${\bf h_k w_j}=0$ for $k \neq j$. The zero-interference can be obtained using the pseudoinverse of the channel gain matrix as weights, so that the received signal can be written as \cite{1603708} \vspace{-0.2cm}
\begin{equation}\label{eq:received_signal}
\begin{split}
y_k&=\left(\sqrt{P_k} {\bf h_k w_k} \right)s_k\\&+\sum_{j\neq k}\left(\sqrt{P_j} {\bf h_k w_j} \right)s_j + z_k,\,\, k=1,\ldots, K
\end{split}
\end{equation}
where $z_k$ is the additive white Gaussian noise at the $k$th user, while $y_k$ is the received signal by the $k$th user. By adopting the zero forcing beamforming (ZFBF), the sum term in \eqref{eq:received_signal} is considered to be zero. Furthermore, for a given D-MIMO network of size $K$, the sum-rate of the ZFBF is given by \cite{1603708} \vspace{-0.4cm}
\begin{equation}\label{eq:sum_rate}
R_{ZFBF} =  \sum_{i=1}^K \log_2\left( 1+P_i\right).
\end{equation}
The optimal $P_i$ can be obtained as $P_i=\left( \mu \gamma_i-1\right)^+$, with ${x}^+$ denoting $\max(x,0)$, and $\mu$ satisfying \vspace{-0.1cm}
\begin{equation}\label{eq:mu}
\sum_{i=1}^K \left( \mu- \frac1{\gamma_i}\right)=KP.
\end{equation}
In \eqref{eq:mu}, $\gamma_i$ is the effective channel gain to the $i$th user \cite{1603708}, defined as $ \gamma_i= 1/||{\bf w_i}||^2$, where $||\cdot||$ denotes the two-norm operation and $P$ represents the power constraint for a single input single output (SISO) transmission.
\vspace{-0.3cm}
\subsection {Frame Structure}
Following the system model in \cite{J:heath_network_partitioning}, the communication is divided into frames of $T$ symbols duration, with each frame consisting of two parts. The first part is devoted to overhead, which includes symbols required for training, feedback, synchronization etc, while the second part is utilized for data transmissions. It has been shown that the length of the first part of the frame is a function of the network MIMO size and equals to $\alpha=\min \left[ \mathcal{L}(K)/T,1\right]$, where $\mathcal{L}(K)$ is the overhead scaling function with respect to the network MIMO size, $K$. The frame duration is assumed to be less than or equal to the channel coherence time (CCT). In our analysis, the effective sum-rate is also defined as the information-theoretic sum-rate (or simply denoted as sum-rate) reduced by the factor of 1-$\alpha$, i.e., $(1-\alpha) R_{ZFBF}$.

\vspace{-0.1cm}
\section{Network Partitioning Optimization}\label{Sec:partioning}
\begin{figure}[t!]
\centering
\includegraphics[keepaspectratio,width=7.5cm]{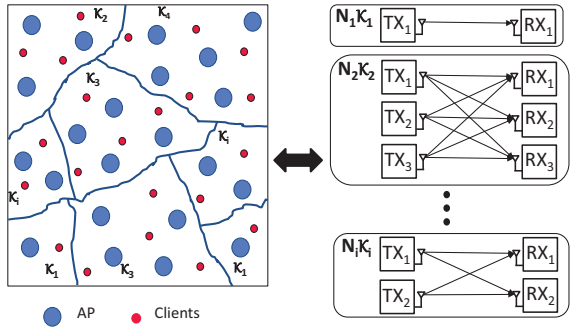}
\caption{Network MIMO is partitioned into orthogonal groups, each with a reduced MIMO size and hence reduced overhead.} \label{Fig:Fig1}
\end{figure}

Towards maximizing the effective sum-rate, the partitioning of the D-MIMO network into orthogonal (e.g., in terms of TDMA) JMB groups, each with a reduced MIMO size, has been proved a promising approach (see Fig.~\ref{Fig:Fig1}). Using orthogonal groups decreases the MIMO size, and hence the required overhead per partition, but also decreases the spectrum utilization, resulting in an interesting trade-off.
Considering a network with $D$ partitions and $\mathcal{K}_{d}$\footnote{For conciseness of presentation, subscript $d$ will denote the MIMO size, e.g., $R_{ZFBF,d}$ will denote the sum-rate of $d$th size MIMO with ZFBF.} APs and clients, the overhead scaling function is an exponentially increasing function equal to $\mathcal{L}(\mathcal{K}_{d})=\mathcal{K}_{d}^r$, where $r>0$ denotes the exponential growth.  More specifically, following the approach proposed in \cite{C:Thukral,J:heath_network_partitioning}, the clients send $F_d$ bits to all the APs, in order to construct the corresponding beamforming weights {$\bf w_i$}. Hence, a total of $KF_d$ is broadcast by each client and thus each AP receives a total of $K^2F_d$ feedback bits from all the clients (i.e., for the case considered in this paper $r=2$).

\vspace{-0.3cm}
\subsection{Optimal Partitioning}
Considering the overhead model in \cite{J:heath_network_partitioning} and clients that are partitioned in $D$ index sets, $1\leq $D$ \leq K$ with $\mathcal{K}_d$ clients in the $d$th group, the optimal solution aims to find the partitioning combination that maximizes the total effective sum-rate, i.e., \vspace{-0.2cm}
\begin{equation}\label{knap1}
\text{\emph{maximize }}  \sum_{d=1}^{K} N_d\, \overline{\alpha}_d \underset{ }{R_{ZFBF,d}}
\text{\emph{ subject to    }}  \sum_{d=1}^{K} N_d \mathcal{K}_{d}=K.
\end{equation}
In \eqref{knap1}, $\overline{\alpha}_d$ and $\alpha_d$ represent the portion of the frame of the $d$th type partition used only for data and overhead, respectively, $\left( i.e., \overline{\alpha}_d=1-\alpha_d=\frac{1}{D}-\frac{\mathcal{K}_d^r}{T} \right)$ and $N_d$ represents the total number of $d$th type partitions that have been generated, with $1 \leq N_d \leq K$. For example, when $K=4$, the optimization in (\ref{knap1}) will aim to find the maximum effective sum-rate among the partitioning combinations [(4x4), (3x3, 1x1), (2x2, 2x2), (2x2, 1x1, 1x1), (1x1, 1x1, 1x1, 1x1)]. We solve this problem by exhaustive search among these combinations. The results serve as a benchmark for the suboptimal overhead constrained partitioning, presented in the next subsection.

\vspace{-0.1cm}
\subsection{Constrained Partitioning and the Knapsack Algorithm}\label{sub_knapsack}
In this section, we investigate the more realistic case, where a maximum allowed overhead size is considered. \subsubsection{Initial Problem} The overhead portion of the frame, is less than or equal to a predefined threshold $\alpha_{\rm th}$, i.e., $\sum_{d=1}^{K} N_d \alpha_d\leq\alpha_{\rm th}$ (with $\alpha_{\rm th} \in [0,1]$), yielding the following optimization problem \vspace{-0.2cm}
\begin{equation}\label{eq:BKP_constraint}
\begin{split}
\text{\emph{maximize }} & \sum_{d=1}^{K} N_d \, \overline{\alpha}_d \underset{ }{R_{ZFBF,d}}  \\
\text{\emph{subject to }}& \sum_{d=1}^{K} N_d \mathcal{K}_{d}=K \, \text{\rm  and } \sum_{d=1}^{K}N_d\alpha_d \leq \alpha_{\rm th}.
\end{split}
\end{equation}%
The optimization problem appearing in \eqref{eq:BKP_constraint} is known as the bounded Knapsack problem (BKP) \cite{B:knapsack}, where $\underset{ }{R_{ZFBF,d}}$ denotes the profit, $\alpha_d$ the weight and $\alpha_{\rm th}$ the capacity of the optimization problem. In this problem the target is the selection of a number $N_d \,(d=1,\ldots,K)$ of partitions of each type, so as to maximize the effective throughput, subject to a maximum allowed overhead size, i.e., $\sum_{d=1}^{K}N_d\alpha_d \leq \alpha_{\rm th}$. \subsubsection{BKP Transformation} The BKP is a generalized formulation of the zero-one Knapsack problem and can be simplified to the latter using the Algorithms $1$ and $2$. Following Algorithm $1$, our BKP is transformed to \vspace{-0.3cm}
\begin{equation}\label{eq:8}
\begin{split}
\text{\emph{maximize }} \sum_{i=1}^{\hat{n}} &\hat{p}_iy_i \\
\text{\emph{subject to }}\sum_{i=1}^{\hat{n}} &\hat{w}_iy_i \leq \alpha_{\rm th} \,\, {\rm and}\,\, \sum_{i=1}^{\hat{n}} \hat{m}_i y_i=K
\end{split}
\end{equation}
where $\hat{p}_i$ and $\hat{w}_i$ represent the sum-rate and the overhead of the $i$th MIMO basic element, respectively, $\hat{n}$ is the total number of the MIMO basic elements, $y_i \in[0,1]$ and $\hat{m}_i$ is the number of AP in the $i$th MIMO basic element.

\begin{table}[t!]
\renewcommand{\arraystretch}{1.1}
\label{Tab:1} \centering
\begin{tabular}{l  }
  \hline
    {\bf Algorithm 1:} BKP to Zero-One Transformation
    \\
    \hline
 {\bf input:} \text{the total number of APs, $n=K$, } \\
 \text{the sum-rate of the $j$th partition, $p_j=R_{ZFBF,j}$, $b_j=\lfloor \frac{K}{j} \rfloor^1$,}   \\
 \text{the overhead factor ($y$th order MIMO, $x$th partition), } $\hat{w}_{x,y}\null=\frac1{x}-\frac{y^r}{T}$
 \\
$\hat{n}:=0;\,\overline{v}:=1$
 \\
{\scriptsize 1:} {\bf for} $j:=1$ to $n$ { \bf do} \  \ \ \ \  \ \text{\% Find the different basic MIMO elements}
\\
{\scriptsize 2:} $k:=0$
\\
 {\scriptsize 3:}{ \bf repeat}
 \\
 {\scriptsize 4:} $\hat{n}:=\hat{n}+1$
 \\
 {\scriptsize 5:} $\hat{p}_{\hat{n}}:=(k+1)p_j$ \  \ \ \ \ \ \ \text{\% Sum-rate of $\hat{n}$th basic MIMO elements}
\\
  {\scriptsize 6:} $k:=k+1$
 \\
 {\scriptsize 7:} $\hat{m}_{\hat{n}}:=kj$  \  \ \ \ \ \ \ \text{\% Number of APs in $\hat{n}$th basic MIMO elements}
 \\
{\scriptsize 8:}$\hat{q}_{\hat{n}}:=k$
\\
{\scriptsize 9:} $\hat{r}_{\hat{n}}:=j$
\\
{\scriptsize 10:} {\bf until} $k=b_j$
\\
{\scriptsize 11:} {\bf end}
\\
\hline
$\null^1 \lfloor \bullet \rfloor$ is the floor function.\\
\end{tabular}
\end{table}

\begin{table}[t!]
\renewcommand{\arraystretch}{1.1}
\label{Tab:2} \centering
\begin{tabular}{  l}
  \hline
    {\bf Algorithm 2:}  Sum-Rates and Overhead of Basic MIMO Elements \\
    \hline
   $I:=\left\{ I_1, I_2, \ldots, I_K \right\}$, with $I_i \in \mathbb{N}^*$, $I_i \neq I_j \, \forall\, i,j \in K, \,\, I_i \leq K$
 \\
 \text{\% Find the maximum number of different partitions} \\
 $R:=\max{\underset{I_i \neq I_j}{\Big| {\rm sum} \left(I_1, I_2, \ldots, I_K \right)\Big|}}=K$
 \\
 $i:=1$
\\
{\scriptsize 1:} {\bf repeat}
\\
 \text{\% For all previously derived basic MIMO elements} \\
{\scriptsize 2:}\,\, {\bf for} $\ell_1, \ell_2, \ldots, \ell_i:=1$ to $\hat{n}$ { \bf do}
 \\
   \text{\% Check if the combination is a valid network partition}\\
{\scriptsize 3:}\,\, { \bf if} ${\underset{\ell_1\neq\ell_2\neq \cdots \neq \ell_i}{\hat{m}_{\ell_1}+\hat{m}_{\ell_2}+\cdots+\hat{m}_{\ell_i}=K}}$
 \\

{\scriptsize 4:}\,\,$\overline{p}_{\overline{v}}:=\sum_{q=1}^i\hat{w}_{\lambda\null^2, \hat{r}_{\ell_q}}\hat{p}_{\ell_q}$ \ \ \ \  \text{\% Sum-rate of the $\overline{v}$th partition } \\
 {\scriptsize 5:}\,\,$\overline{w}_{\overline{v}}:=\hat{w}_{\lambda, \hat{r}_{\ell_1}}+\sum_{q=2}^i \hat{q}_{\ell_q}\hat{w}_{\lambda, \hat{r}_{\ell_q}}$
\text{\% Overhead of the $\overline{v}$th partition } \\
   {\scriptsize 6:}\,\,$\overline{v}:=\overline{v}+1$
 \\
{\scriptsize 7:}\,\,{\bf end}
\\
{\scriptsize 8:}\,\,{\bf end}
\\
{\scriptsize 9:}\,\, $i:=i+1$
\\
{\scriptsize 10:} { \bf until} $i=R+1$
\\
 {\bf output:} $\overline{v},  \overline{p}_i,\overline{w}_{i}$
\\
\hline
$\null^2 \lambda=\sum_{h=1}^i \ell_h$. $ \mathbb{N}^*$ denotes the set of natural numbers excluding zero.   \vspace{-0.2cm}
\end{tabular}
\end{table}
The functionality of Algorithm $1$ can be better explained with an example as follows. Assuming $K=4$, using Algorithm $1$ and \eqref{eq:8}, BKP is transformed to the following problem: find the network partitioning combination that maximizes the effective sum-rate, using $8$ different MIMO basic elements, namely [1x1, $2*(1$x1), $3*(1$x1), $4*(1$x1), 2x2, $2*(2$x2), 3x3, 4x4], (each characterized by a different sum-rate and overhead, computed using Algorithm $1$) subject to a predefined threshold for the maximum allowed overhead.

The second step for transforming BKP to a zero-one Knapsack problem is made by employing Algorithm $2$. In this algorithm all potential combinations of the MIMO basic elements are determined (taking into account the total number of APs in the network). Specifically, the outcomes of this algorithm are:
 \begin{itemize}
   \item the total number of different partitioning combinations $\overline{v}$ [e.g., for $K=4$, $\overline{v}=5$, i.e., (4x4), (3x3,1x1), $2*(2$x2), (2x2, $2*(1$x1)), $4*(1$x1)]
   \item the corresponding performances in terms of the effective sum-rate and overhead $\overline{p}_i, \overline{w}_i$, respectively, with $i \in [1,\overline{v}]$.
 \end{itemize}
\subsubsection{Final Knapsack Solution}Following these two algorithms, the initial BKP has been finally transformed to the following simplified zero-one Knapsack problem \vspace{-0.3cm}
\begin{equation}
\begin{split}
\text{\emph{maximize }}\sum_{i=1}^{\overline{v}} \overline{p}_iz_i
\text{\emph{ \ \ subject to \ \ }}\sum_{i=1}^{\overline{v}} \overline{w}_iz_i \leq \alpha_{\rm th}
\end{split}
\end{equation}
where $z_i \in[0,1]$. The exact optimal solution to this zero-one Knapsack problem is a straight-forward procedure and it coincides with the one provided via the Greedy-Split algorithm. Specifically, the following steps are followed:
\begin{itemize}
\item $\overline{p}_i$ $(i \in \overline{v})$ are sorted on descending order, i.e., $\overline{p}_1 =\max (\overline{p}_i)$
\item examining $\overline{p}_i , \overline{w}_i$ in ascending order
\item if \hspace{0.3cm}$\overline{w}_i\leq\alpha_{\rm th}$
\item  solution$\rightarrow \overline{p}_i, \overline{w}_i$.
\end{itemize}
The maximum number of searches is proportional to the number of different possible partitions, given by the \emph{partition function} \cite[\S24.2.2]{B:abramowitz} (e.g., 42 for a $10{\rm x}10$ D-MIMO system) hence the complexity of the proposed approach is low.

\begin{figure}[b!]
\centering
\includegraphics[keepaspectratio,width=5.5cm, trim=5cm 1.0cm 4cm 3.0cm]{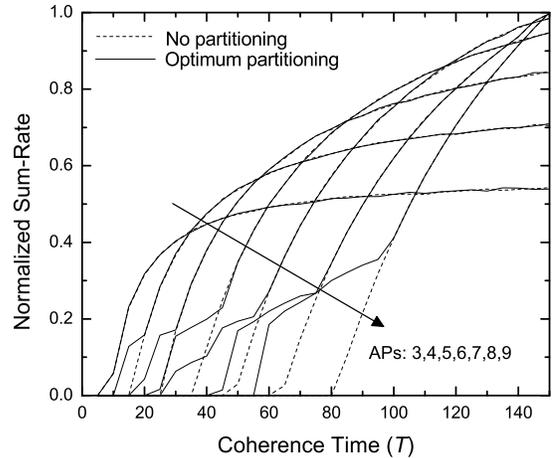}
\caption{The normalized sum-rate (NSR) with and without network partitioning. The sum-rate is normalized by the corresponding performance of the optimal partitioning of a 9{\rm x}9 MIMO network with SNR=25dB.} \label{Fig:Fig2}
\end{figure}
\begin{figure}[t!]
\includegraphics[keepaspectratio,width=6.5cm, trim=1cm 1.2cm 8cm 3.5cm]{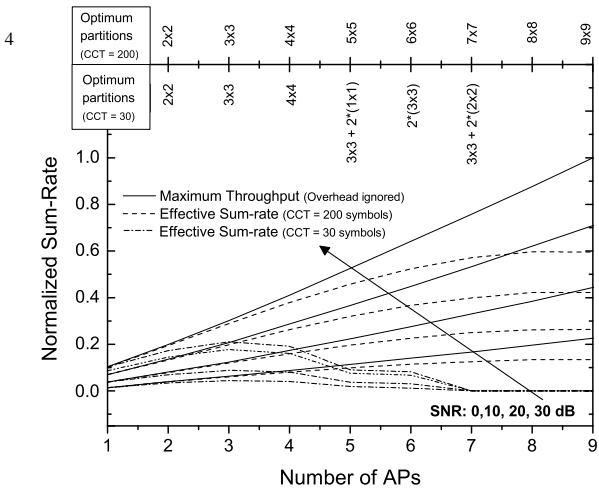}
\caption{The maximum achievable NSR (ideal case) and the effective NSR with optimal partitioning vs the number of APs. The sum-rates are normalized by that achieved by the $9{\rm x}9$ scheme for SNR = 30dB and zero overhead.} \label{Fig:Fig3}
\end{figure}

\begin{figure}[t!]
\centering
\includegraphics[keepaspectratio,width=7cm,trim=4cm 0.5cm 4cm 1.5cm]{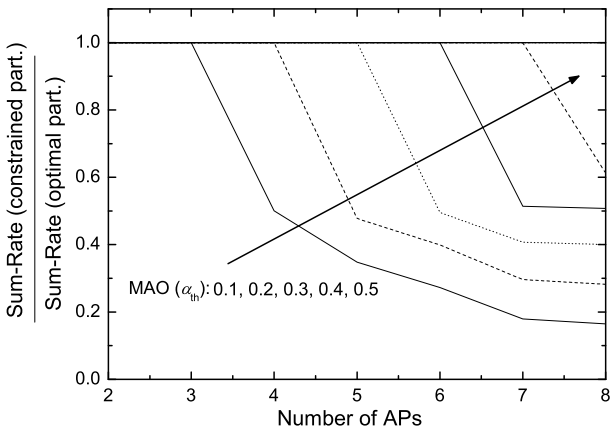}
\caption{The suboptimal partitioning sum-rate, as a percentage of the sum-rate achieved with the optimal partitioning vs MAO.} \label{Fig:Fig5}
\end{figure}

\vspace{-0.3cm}
\section{Performance Results and Discussion}\label{sec:numerical_results}
For the performance evaluation results presented in this section, the parameters considered in all cases are: randomly deployed single-antenna APs and clients, independent and identically distributed (i.i.d.) Rayleigh fading channels \footnote{Assuming non i.i.d. fading greatly increases the searching
complexity, since the possible partitioning combinations significantly increase. Generating
the channels i.i.d. gives insight into the best possible performance \cite{J:heath_network_partitioning}.}, while for TDMA, overhead is assumed to scale linearly with the number of users. The benefits of partitioning a D-MIMO network into orthogonal groups, each employing multi-user beamforming, are illustrated in Fig.~\ref{Fig:Fig2}, where the normalized sum-rate (NSR) is plotted as a function of the CCT. It can be observed that partitioning improves the effective sum-rate of the network as the coherence time decreases, while the NSR gains become more pronounced for larger MIMO networks. As the coherence time increases employing multi-user beamforming for the whole MIMO network becomes more efficient.

In Fig.~\ref{Fig:Fig3}, the NSR is plotted as a function of the number of the APs for various values of SNR and CCT. The NSR has been evaluated for both cases of maximum achievable sum-rate (ideal case without considering the overhead) and effective sum-rate (with optimal partitioning). It is depicted that the performance improves as the SNR or the number of APs increases. For the effective NSR case, it is important to note that the linear scaling of the sum-rate is not maintained as the number of APs increases, due to the overhead, whilst the performance considerably improves for high values of CCT. It should be also mentioned that the partition types change as the CCT increases, as it is depicted at the top axis of this figure. For example, the notation $3{\rm x}3+2*(1{\rm x}1)$ denotes that the $5{\rm x}5$ MIMO network is partitioned into one partition employing a $3{\rm x}3$ MIMO and $2$ partitions with a $1{\rm x}1$ SISO scheme.

Finally, Fig.~\ref{Fig:Fig5} depicts the sum-rate of the constrained suboptimal partitioning as a percentage of the corresponding one with optimal partitioning, as a function of the number of APs and for various values of the maximum allowed overhead (MAO). In this figure, the optimal and suboptimal partitioning schemes proposed in Section~\ref{Sec:partioning} are compared, in order to highlight the performance degradation due to the constrained overhead length within the frame. It is depicted that for low values of the MAO, the overhead-constrained sum-rate is relatively small compared to the unconstrained one. As MAO increases, the relative performance between these two metrics also increases. Finally, it is noted that for small networks dimensions, the performance of the constrained sum-rate is approaching that of the optimal partitioning.

\vspace{-0.1cm}
\section{Conclusions}\label{sec:conclusions}
In this letter, the optimal effective sum-rate of D-MIMO is investigated for the cases where the overhead subframe size can either change dynamically or must be fixed. For the first approach, exhaustive search is employed, while for the second one, the partitioning problem is formulated for the first time as an elegant Knapsack optimization problem. Performance evaluation results show that linear scaling of the effective sum-rate is not always maintained as the number of APs increases, due to the overhead required. Finally, relaxing the assumptions for the orthogonality among the partitions opens up an interesting investigation field that will be included in our future research activities.
\vspace{-0.3cm}


\vspace{-0.cm}
\bibliographystyle{IEEEtran}
\bibliography{IEEEabrv,Lioumpas_WCL2013-0449.R1}
\end{document}